 \documentstyle[aps,epsfig]{revtex}
\newcommand{\eq}{\begin{equation}} 
\newcommand{\eqx}{\end{equation}}
\newcommand{\eqn}{\begin{eqnarray}} 
\newcommand{\eqnx}{\end{eqnarray}}
\newcommand{\PO}{\rm l \! P }
\newcommand{\RO}{\rm l \! R }

\begin{document}
 
\title{Hard diffraction and  the nature of the Pomeron}
\author{J. Lamouroux\thanks{Universit\'e Joseph% 
Fourier, BP53 - 38041 Grenoble CEDEX 9, France}, R. Peschanski\thanks{%
CEA/DSM/SPhT,  Unit\'e de recherche associ\'ee au CNRS, CE-Saclay, F-91191 
Gif-sur-Yvette Cedex,
France}, C. Royon\thanks{%
CEA/DSM/DAPNIA/SPP, F-91191 
Gif-sur-Yvette Cedex,
France} and L. Schoeffel\thanks{{\it idem }}}
\maketitle

\begin{abstract}
We ask the question whether  the quark and gluon distributions in the Pomeron 
obtained from QCD fits to hard diffraction processes at HERA can be 
dynamically generated  from a state made of {\it valence-like}  gluons and sea 
quarks as input. By a method combining {\it backward}   $Q^2$-evolution for data 
exploration  and {\it forward}  $Q^2$-evolution for a best fit determination, we 
find that  the diffractive structure functions  published by the H1
collaboration at HERA  can be  described by a  simple {\it 
valence-like} input at an initial scale of order  $\mu^2 \sim 2.3-2.7\ GeV^2$.  
The parton number sum rules at the initial scale  $\mu^2$ for the H1 fit gives 
$2.1\pm .1\pm .1$ and 
$.13\pm .01 \pm .02$ for gluon and sea quarks respectively, corresponding to  an 
initial Pomeron  state made of  (almost) only two gluons. It has    flat 
gluon  density leading to a  plausible interpretation in terms of a 
gluonium state.
\end{abstract}
\bigskip
{\bf 1. Introduction}
\bigskip

Since years, the Pomeron remains a subject of many  interrogations. Indeed, 
defined as the virtual colourless carrier of strong interactions,  the nature of 
the Pomeron is still a real challenge. While in the perturbative 
regime  of QCD it can be defined  as a compound system of two Reggeized gluons 
\cite {bfkl} in the approximation of resumming the leading logs in  energy, its 
non-perturbative structure is basically unknown.

However, in the recent years, an interesting experimental  investigation on 
``hard'' diffractive processes led to a new insight into  Pomeron problems. At 
the HERA accelerator, it has been discovered that  a non negligeable 
amount of $\gamma^*\!-\!{\rm proton}$ deep inelastic events can be produced with 
no 
visible breaking of the incident proton. There are various phenomenological 
interpretations of this phenomenon, but a very appealing one (which indeed 
constituted a prediction \cite {ingelman}) relies upon a partonic interpretation 
of  the structure of the Pomeron. In fact, it is possible to nicely describe the 
two sets of cross-section data from H1 \cite{f2dH1} and ZEUS \cite{f2dZEUS} 
(after taking into account also a Reggeon component) by a QCD DGLAP evolution of  
parton distributions in the Pomeron combined with Pomeron flux factors 
describing phenomenologically the probability of finding a Pomeron state in the 
proton.  Sets of  quark and gluon distributions in the Pomeron following 
LO or NLO $Q^2$-evolution equations are obtained \cite{us}, which 
successfully describe 
H1 and ZEUS data sets separately\footnote{Both sets are compatible within the 
errors but  a difference was observed between the two sets of structure 
functions corresponding to different trends in the  $Q^2$-dependence of the data 
\cite{us}. This motivates the separation between the two sets of data for the 
analysis.}. 

The idea carried on in the present work is to find whether the Pomeron structure 
functions can be obtained from a standard DGLAP \cite {dglap} evolution 
equation initialized by a {\it valence-like} input at some scale $\mu.$ By 
{\it valence-like}  we mean an input distribution at low scale $\mu^2$ for which 
both the density  of gluons and sea quarks remains  finite when  $\beta \to 0,$ 
where $\beta$ is  the energy-momentum shared between 
the constituants. If this is achieved, the number of gluons and sea quarks in 
the Pomeron is well-defined and finite. This may give an information on the 
non-perturbative origin of the Pomeron which can then be interpreted as a state 
made of constituent gluons and sea quarks with a given $\beta$ 
distribution (which reflects the energy-momentum sharing between 
the constituants and thus their interaction) and a given transverse size given 
by the scale $\mu.$ 

Our approach is inspired by the well-known GRV approach for the proton 
\cite{GRV} (as well as its extensions to  pion, photon) where the gluon, quark 
and antiquark distribution are obtained {\it via } dynamical parton generation 
from LO or NLO DGLAP  evolution starting at low scale. In the case of the 
proton, taken as an example, the idea was in principle  to  generate all proton 
distributions from QCD radiation starting only with the three valence quarks. 
After a series of  refinements due to precise data analysis \cite{GRV}, it 
happens   necessary to introduce also {\it valence-like} antiquarks and gluons. 
In 
any case,   the overall picture leads to the useful GRV set of structure 
functions  widely  used in QCD phenomenology.

The main results of our analysis are herafter summarized:

{\bf i)} The H1 set of data is fully compatible with an initial 
{\it valence-like} set of sea $(S \equiv q+\bar q)$ and gluon $(G)$ 
distributions 
at a scale $\mu^2_{NLO} \sim 2.7\ GeV^2$ ($\mu^2_{LO} \sim 2.3\ GeV^2$).
 
{\bf ii)} Simple input {\it valence-like} distributions 
\begin{eqnarray}
&~& \beta S (\beta, Q^2= \mu^2) = a_q\ \beta\ (1 - \beta)^{\ b_q} \nonumber\\
&~& \beta G (\beta, Q^2= \mu^2) = a_g \  \beta\ (1 - \beta)^{\ b_g} \ ,
\label{valence}
\end{eqnarray}
give rise to a good fit of data. The four  parameters $a_{q,g},b_{q,g}$ are 
given in table I with 
both statistical and systematic errors.

{\bf iii)}. The values obtained for $b_g,b_q$ are small, leading to flat initial 
parton density distributions for the Pomeron at the initial scale $\mu^2.$ 
The exponential term in the usual parametrizations of input quark and gluon 
densities \cite{f2dH1,us}, with an essential singularity at $\beta=1,$ is needed 
to    compensate for the singularities of QCD matrix elements at low $Q^2.$  
This explains the  difference between the  usual initial parametrizations for 
diffractive {\it vs.}  total proton structure functions.

{\bf iv)} The ZEUS set of data, when parametrized using (\ref{valence}), leads 
to a worse $\chi^2$ fit, see table I. Even if these data have smaller 
weight in a global analysis, this difference deserves further study.

{\bf v)} Considering the  obtained set of Pomeron structure functions, the 
input parton distributions of the Pomeron can be interpreted as those of a 
gluonium in a fundamental quantum state. Indeed, the gluon and sea quark number 
QCD sum rules (see formula (\ref{numbers}) in section {\bf 4}), correspond to a 
state
of (almost only) two interacting gluons with  (almost) flat parton density in 
$\beta.$

\bigskip\bigskip

 {\bf 2. Backward evolution}

\bigskip
Our method can be decomposed in two steps. First, we consider the existing 
parametrizations of the Pomeron structure functions, which are not of GRV type. 
We perform a {\it backward} QCD evolution in order to see whether at the same 
small 
$Q^2$ scale, the sea quark and gluon distributions can be compatible with the 
{\it 
valence-like} property. Then, if this investigation leads to a positive result, 
we perform a new QCD analysis of data, starting directly from  a {\it 
valence-like} input. A consistency check is to verify that the new 
parametrizations of the Pomeron structure functions are compatible, within 
errors, with the initial ones.

 As an illustration of our method for generating  dynamical parton 
distributions, 
let us first consider the {\it backward} evolution for the proton within the 
 GRV scheme \cite{GRV}. By construction, the parton distributions are obtained 
from {\it valence-like}  quark, antiquark and gluon distributions at a small 
scale 
$\mu^2.$ In the Mellin $j$-plane, the {\it valence-like} inputs correspond to 
Mellin transformed moments, or more generally continuous  $j$-distributions, 
which remain finite when $j\!\to \!1.$ Indeed, the $j\!=\!1$ moment, when it is 
well-defined and normalized to the $j=2$ energy-momentum sum rule, defines the 
number of partons. It is constant for valence quarks for any $Q^2,$ while it is 
in general infinite for sea quarks and gluons, except eventually at the input 
scale $\mu^2,$ if and only if the input distributions are {\it valence-like}. 
 
\begin{figure}
\begin{center}
\centerline{\psfig{figure=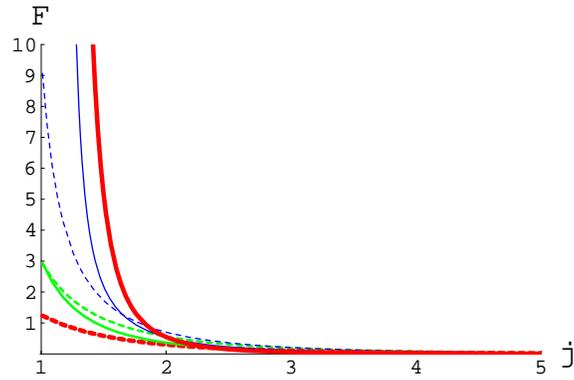,height=2.in}}
\end{center}
\bigskip
\caption{{\it Backward $Q^2$-evolution of the proton $j$-distributions from 
GRV.} 
The backward NLO evolution of the proton $j$-distributions calculated 
from 
the GRV 
parametrizations is shown at the initial scale  $\mu^2_{NLO}=.4\ GeV^2$ 
(dashed lines) and at 
$Q^2=12\ GeV^2$ (continuous lines). Dark lines: Sea 
quarks; Dark grey lines: gluons; light grey lines: valence quarks.
}
\label{fig1}
\end{figure}

In the case of  the total structure functions of the proton, this is what can be 
seen in 
Fig.\ref{fig1}, where we have reproduced the NLO {\it backward} 
evolution 
back to $\mu^2$ of the valence, sea  quark and gluon distributions from  the GRV 
parametrization. It shows that 
all 
distributions which are infinite at  $Q>\mu$ become finite at  $j\!\to \!1$ when 
$Q\!\to \!\mu.$ In this instructive exercise we use the Mellin transform 
formalism for the  (LO and NLO)  
$Q^2$-evolution\cite{lacaze}, which is  particularly 
suitable for our purpose since DGLAP equations take a simple two-by-two matrix 
form in Mellin space and allow us to use an  exact  analytical solution for the 
Pomeron structure functions at all $Q^2.$ We use the same method for the 
diffractive structure functions.
 
In our search for a {\it valence-like} input for the Pomeron, we 
now use the NLO {\it backward} evolution starting from the ans\"atze of 
Ref.\cite{us}. Indeed, the key technical point of our analysis is  the 
identification of an exact analytic  Mellin transform of the input 
parametrizations commonly used \cite{f2dH1,us} for the DGLAP evolution of the 
Pomeron structure functions.

Starting from the parametrizations at $Q_0^2=3\;GeV^2$ \cite{us}, the  sea quark 
distribution $ 
{S}(\beta,Q^2)=\sum_{flavors}(q+\bar q)(\beta,Q^2)$
and the gluon distribution $ {G}(\beta,Q^2)$ are parameterized in 
terms of coefficients $C_j^{(G,S)}$ : 
\eq
\beta{\it {S}}(\beta,Q_0^2) = \left[
\sum_{j=1}^n C_j^{(S)} \ P_j(2\beta-1) \right]^2
\ e^{\frac{a}{\beta-1}} \;\ ; 
\ \beta{\it {G}}(\beta,Q_0^2) = \left[
\sum_{j=1}^n C_j^{(G)} \ P_j(2\beta-1) \right]^2
\ e^{\frac{a}{\beta-1}}\ ,
\label{gluon}
\eqx
where 
$P_j(\zeta)$ is the
$j^{th}$ member in a set of Chebyshev polynomials, 
$P_1=1$, $P_2=\zeta$ and $P_{j+1}(\zeta)=2\zeta P_{j}(\zeta)-P_{j-1}(\zeta).$ 
The parameters $C_j^{(G,S)}$ used in \cite{us} are given in Table II. 
In formula (\ref{gluon}), one takes the parameter $a=.01\ .$

Now, let the structure functions be expanded  as 
$\sum_{i=0}^{4}\;d_{i}^{(G,S)}\;\beta^i\;\exp 
[a/(\beta-1)],$ with straightforward linear relations between  $d_i^{(G,S)}$ and  
$C_i^{(G,S)};$ 
The Mellin transform reads \cite{grad}:
\begin{equation}
F^{(G,S)}(j,Q^2) \equiv 
\sum_{i=0}^{4}\;d_{i}^{(G,S)}\;\Gamma(i\!+\!j\!-\!1)\ e^{-\frac{a}{2}
}
\;W_{1\!-\!i\!-\!j,\frac 12}(a)\ ,
\label {Mellin}
\end{equation}
where the Whittaker function $W_{1\!-\! i\!-\! j,\frac 12}(a)$ can also be 
expressed in terms of the Meijer function $G_{24}^{40}$  \cite{grad}.

Once using expressions (\ref{Mellin}), together with the NLO or LO evolution 
scheme
\cite{lacaze}, it is straightforward to get the whole $j$-dependent parton 
distributions in any suitable range of $Q^2,$ either for {\it forward} $(Q>Q_0)$ 
or {\it backward} $(Q<Q_0)$ evolution.

For the Pomeron, looking for a {\it valence-like} input, we  use the NLO {\it 
backward} evolution starting from the ans\"atze of Ref.\cite{us} and look for 
the possibility of {\it valence-like} distributions in some range of $Q^2 \sim 
\mu^2.$ The 
analytic singularities of the Whittaker function in (\ref{Mellin}) are 
approximately
 cancelled by those of the  evolution matrix elements. They are both 
situated at $j=1$ and the $Q^2$ {\it backward} evolution induces a  change of 
sign  both for the sea and 
the glue distributions in the same range  $Q^2 \sim \mu^2.$ 

\begin{figure}
\begin{center}
\centerline{\psfig{figure=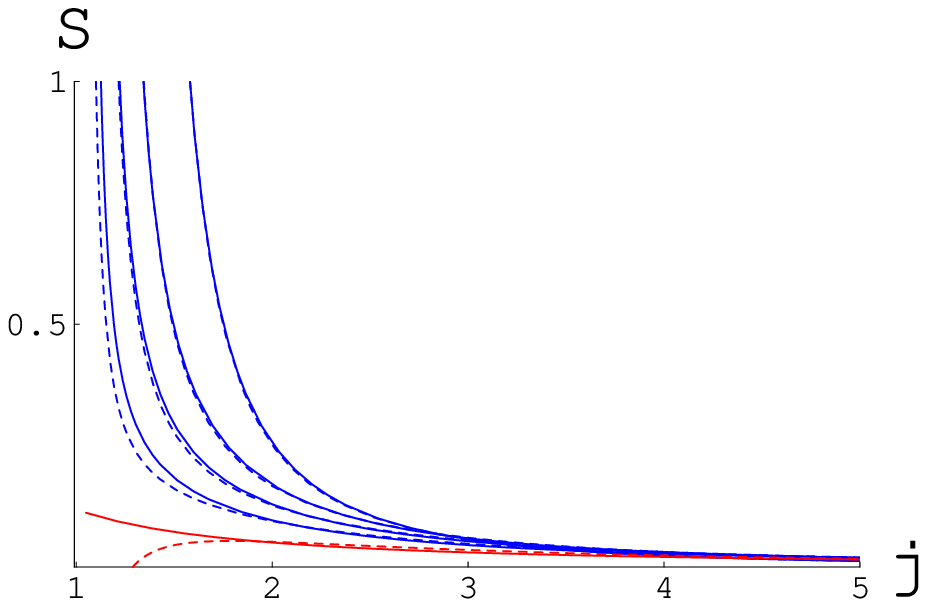,height=1.8in}}
\centerline{\psfig{figure=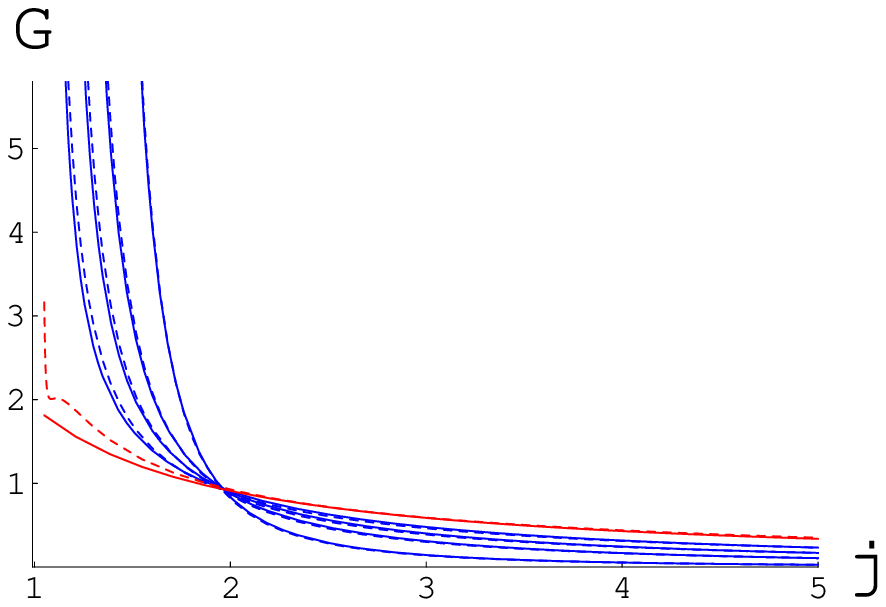,height=1.8in}}
\end{center}
\bigskip
\caption{$Q^2$-evolution of the Pomeron $j$-distributions for the H1 case: 
comparison between {\it backward} and {\it forward evolution}.  Dashed 
lines:  
Backward NLO evolution; 
Continuous lines: exemple of valence-like inputs.  Curves from top to bottom are 
displayed for 
$Q^2=10^6, 75, 12, 4.5\ GeV^2 $ and $\mu^2_{NLO}.$ 
}
\label{fig2}
\end{figure}

\begin{figure}
\begin{center}
\centerline{\psfig{figure=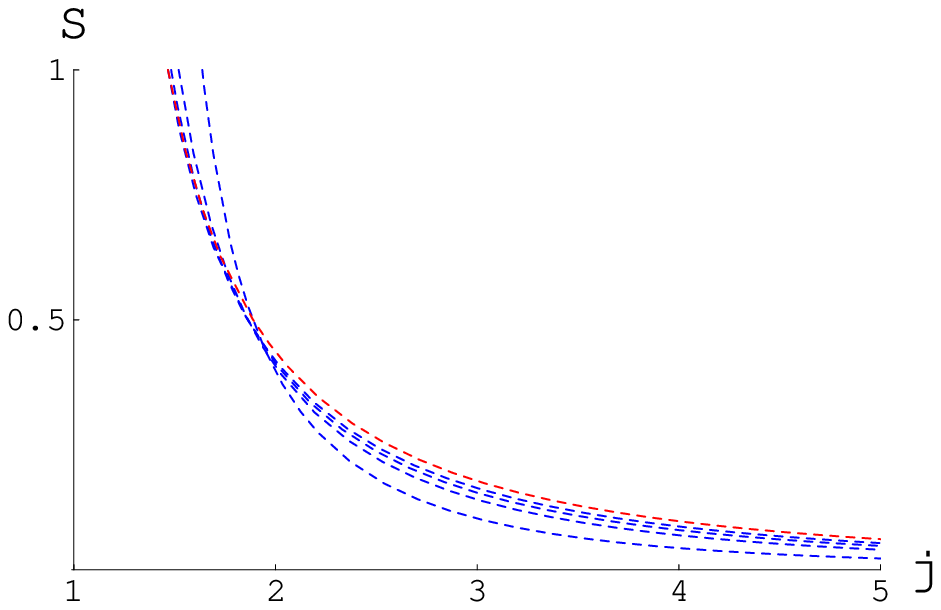,height=1.8in}}
\centerline{\psfig{figure=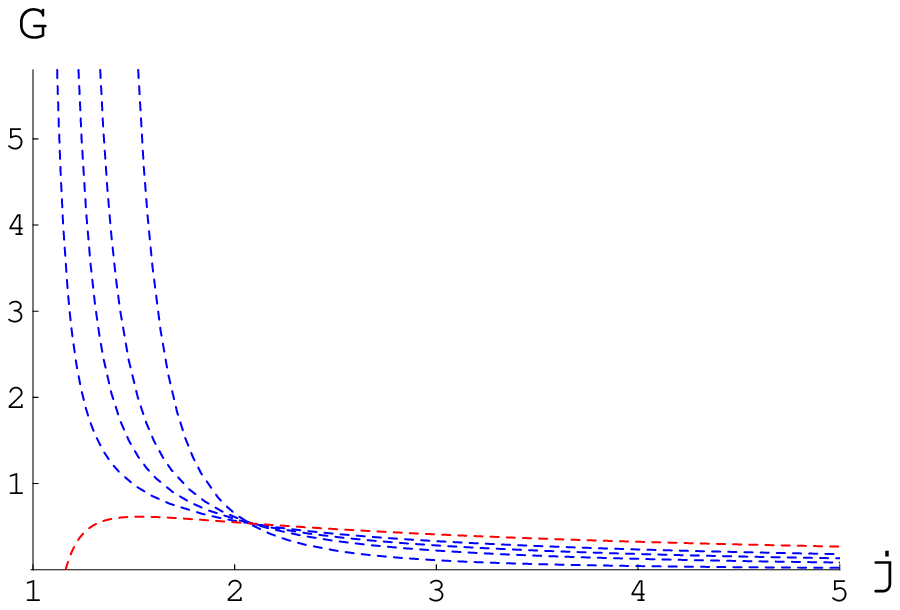,height=1.8in}}
\end{center}
\bigskip
\caption{Backward $Q^2$-evolution of the Pomeron $j$-distributions: ZEUS case. 
The dashed lines correspond to the Backward NLO evolution of the 
Pomeron  
$j$-distributions [5]. Same $Q^2$ values 
as 
in Fig.2.
}
\label{fig3}
\end{figure}

The results  (dashed lines) are shown in Fig.\ref{fig2} for the H1 sea and gluon 
$j$-distributions and in Fig.\ref{fig3} for ZEUS. 
Let us first comment the results obtained for the parametrizations of H1 data. 
As shown in Fig.\ref{fig2}, the behaviour of the Mellin transformed 
distributions show a different  trend for small $Q^2$  values.  The 
singularities present at $j=1$ in the initial parametrizations (\ref{Mellin}) 
interfere negatively with those present in the evolution matrix elements 
\cite{lacaze}. The key point is that they do so  both for  sea and  gluon 
distributions in the same range of  $Q^2 \approx 2\ GeV^2$ below 
which a transition occurs. Around this value, the singularities in matrix 
elements overcome the initial ones. In the same figure, a qualitative  exemple 
of {\it forward} evolution starting from a {\it valence-like} input, showing how 
the $Q^2$ dependent matrix elements conspire to mimic the original 
parametrizations of \cite{us}. This will be studied in detail  and 
quantitatively confirmed in the next section.

Note that the exponential term in formulae (\ref{gluon}), which was necessary to 
describe the H1 diffractive structure function data \cite{f2dH1,us}, has an 
essential singularity at $\beta=1.$ It can now be understood as being induced by 
the  compensation of the singularities of QCD matrix elements at low $Q^2.$  It 
is related to the flatness of the input parton distributions of the 
Pomeron\footnote{The situation is different for the total structure functions. 
The input distributions \cite{GRV} are decreasing like powers when 
$\beta\!\to\!1.$} that we find in our fits of the diffractive structure 
functions.  

Fig.\ref{fig3} deals with the analysis of the  {\it backward} evolution for 
the parametrizations of ZEUS data performed in paper \cite{us} within the  same 
scheme as for H1 data. Quite interestingly, we did not find a range  of $Q^2$  
below which {\it both} sea and gluon $j$-distributions  meet a transition   when 
$j\to 1.$ In fact, while the glue distribution flips down at lower $Q^2,$ the 
sea  keeps its singular trend. This result can be traced back to the difference 
seen  in the  $\beta$-distributions of 
\cite{us}. The gluon distribution is weaker than for H1, while the sea 
distribution remains larger at $\beta \to 0.$ In some sense, the {\it backward} 
evolution is more drastically  driven by the sea than by the glue. This 
qualitative result of  {\it backward} evolution will also be confirmed by the 
subsequent {\it forward} evolution analysis.

\bigskip
{\bf 3. Forward evolution}
\bigskip

It is well-known \cite{f2dH1,us} that the diffractive
structure function $F_2^{D(3)}$, measured from DIS
events with large rapidity gaps can be expressed as a sum of two factorised 
contributions corresponding to a Pomeron and secondary Reggeon trajectories.
\begin{eqnarray}
F_2^{D(3)}(Q^2,\beta,x_{\PO})=
f_{\PO / p} (x_{\PO}) F_2^{\PO} (Q^2,\beta)
+ f_{\RO / p} (x_{\PO}) F_2^{\RO} (Q^2,\beta) \ ,
\label{reggeform}
\end{eqnarray}
where $x_{\PO}$ is the fraction of energy of the proton flowing into the Pomeron 
or Reggeon.
In this parameterisation,
$F_2^{\PO}$ can be interpreted as the structure function of the 
Pomeron and thus be expressed in a conventional way in terms of $
\beta S (\beta, Q^2)$ and $\beta G (\beta, Q^2)$. The same can be said for for 
$F_2^{\RO},$ with the restriction that it 
takes into account various secondary Regge contributions which can hardly be 
separated and whose structure functions are modelized from the pion ones.
The Pomeron and Reggeon fluxes are assumed to follow a Regge behaviour with  
linear
trajectories $\alpha_{\PO,\RO}(t)=\alpha_{\PO,\RO}(0)+\alpha^{'}_{\PO,\RO} t$, 
such
that
\begin{equation}
f_{{\PO},{\RO} / p} (x_{\PO})= \int^{t_{min}}_{t_{cut}} 
\frac{e^{B_{{\PO},{\RO}}t}}
{x_{\PO}^{2 \alpha_{{\PO},{\RO}}(t) -1}} {\rm d} t 
\label{flux}
\end{equation}
where $|t_{min}|$ is the minimum kinematically allowed value of $|t|$ and
$t_{cut}=-1$ GeV$^2$ is the limit of the measurement. 

The H1 and ZEUS diffractive data \cite{f2dH1,f2dZEUS} were 
fitted using the simple formulae (\ref{valence}) for the sea quark and gluon 
distributions, with five free parameters:  $a_q, b_q, a_g,  b_g$ and  $\mu^2,$ 
the initial scale of the forward  evolution.
In order to avoid a region where the diffractive longitudinal structure
function is large, only data with $y < 0.45$ are included in the fit to find the
Pomeron and Reggeon intercepts. Further more, for the Reggeized flux factors,
we take the same values of the Pomeron and Reggeon trajectory intercept as in 
Ref. \cite{us},
namely $\alpha_P(0)=1.20 \pm 0.09$ and $\alpha_R(0)=0.62 \pm 0.03$ for H1,
and $\alpha_P(0)=1.13 \pm 0.04$ for ZEUS (there is no need for Reggeons for ZEUS
data). Only data points with $Q^2 \ge 3$ GeV$^2$, $\beta \le 0.65$, $M_X >2$ 
GeV,
and $y < 0.45$ are included in the QCD fit to avoid large higher twist effects 
and the region that may be most strongly affected by a non-zero value of $R$, 
the ratio
of the longitudinal to the tranverse diffractive structure functions.
The QCD fit was performed both at leading order and next-to-leading order using
the usual DGLAP \cite{dglap} evolution equation.

\begin{figure}
\begin{center}
\centerline{\psfig{figure=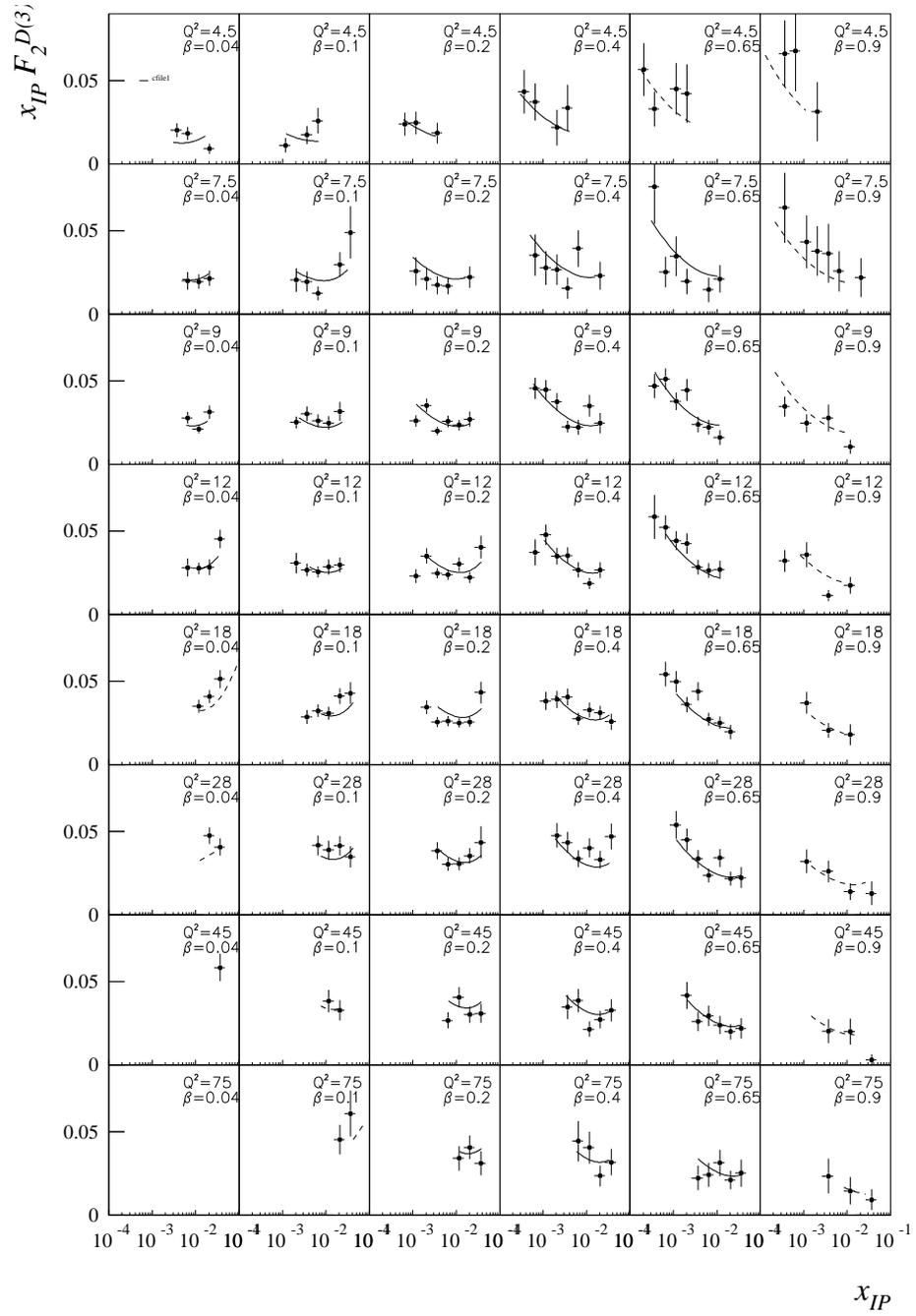,height=7.in}}
\end{center}
\bigskip
\caption{Result of the NLO QCD fit to the H1 $F_2^D$ data. The $\chi2$ is 194.6
/161 data points.
Full line: bins used in the QCD fit, dashed line: extrapolation  to 
the bins
not used in the fit. We note a very good description of the data over the full 
range
using the  simple parametrisation, see formula (\ref{valence}).
}
\label{laurent1}
\end{figure}

The fitted parameter values obtained for the H1 and ZEUS data are given in Table
I. The $\chi^2$ value obtained for the H1 collaboration is quite 
good
($\chi^2$=210.7 and $\chi^2$=194.6
respectively at LO and NLO for 161 data points)
and the fit result at NLO is shown in Fig.\ref{laurent1}.
We  even note a good description of the data at high $\beta$ which are not 
included
in the fit (dashed lines in Fig.\ref{laurent1}).
 
\begin{figure}
\begin{center}
\centerline{\psfig{figure=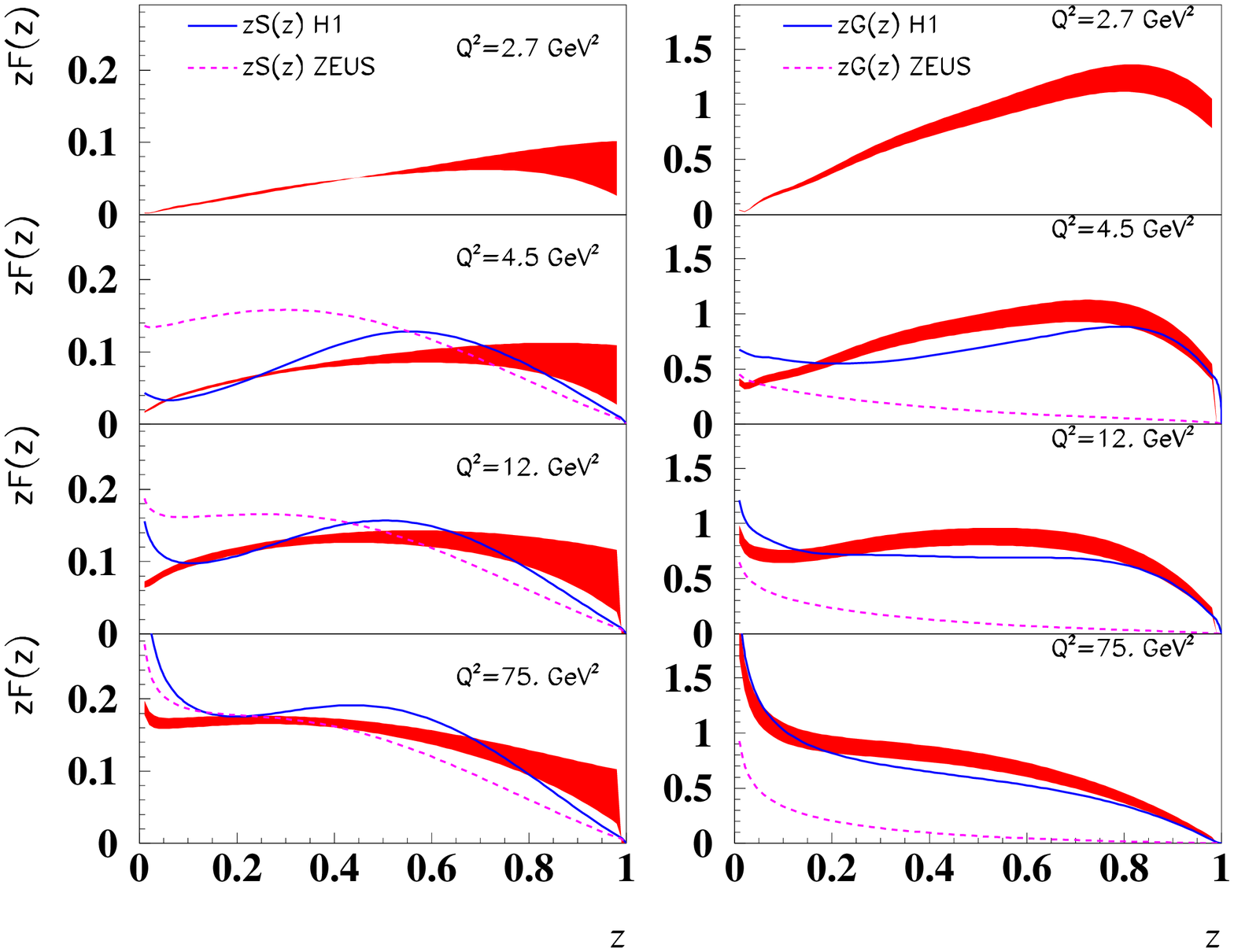,height=5.in}}
\end{center}
\bigskip
\caption{LO  parton strucure functions (grey bands) obtained with the fit to
H1 data. Left: quark density, right: gluon density,
for four different values of $Q^2$, namely 2.7 (the initial scale $\mu^2$), 4.5, 
12.
and 75 GeV$^2$. The error bands correspond to the systematic and statistical 
errors
added in quadrature. The result is compared with the known [5] NLO DGLAP fit in 
full
line for H1 data, and dashed line for ZEUS data. 
}
\label{laurent3}
\end{figure}

\begin{figure}
\begin{center}
\centerline{\psfig{figure=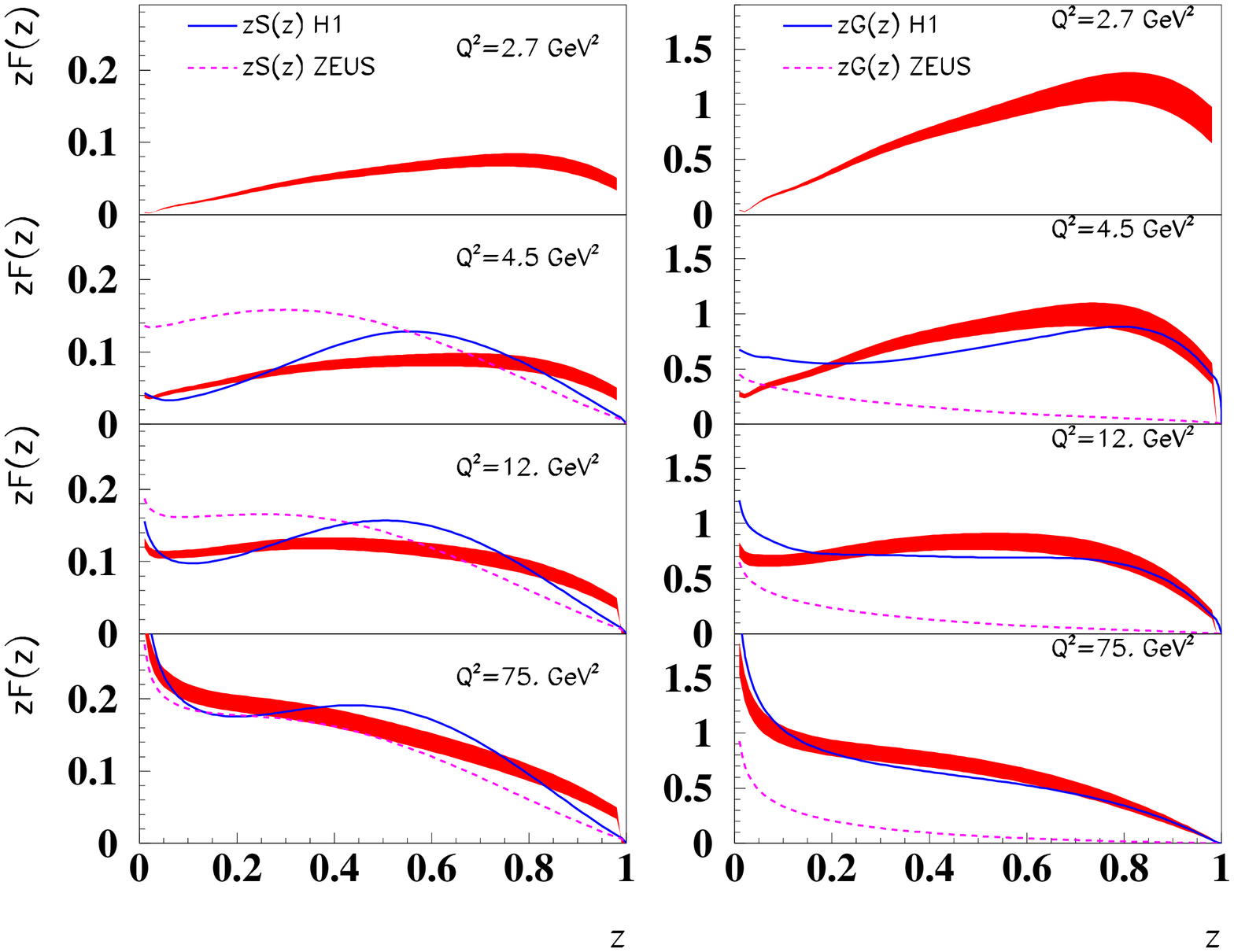,height=5.in}}
\end{center}
\bigskip
\caption{NLO  parton strucure functions (grey bands) obtained with the fit to
H1 data. Left: quark density, right: gluon density,
for four different values of $Q^2$, namely 2.7 (the initial scale $\mu^2$), 4.5, 
12.
and 75 GeV$^2$. The error bands correspond to the systematic and statistical 
errors
added in quadrature. The result is compared with the known [5] NLO DGLAP fit in 
full
line for H1 data, and dashed line for ZEUS data. We note a good agreement 
between
our results and the usual H1 gluon and quark densities.
}
\label{laurent2}
\end{figure}

In Fig.\ref{laurent3} and Fig.\ref{laurent2} are displayed the parton densities 
obtained
with the LO and NLO fits. As expected, we find that the gluon densities is much 
larger than 
the quark one. The input distributions are characterized by a starting scale 
($\mu^2=$2.7 GeV$^2$)  which is higher for diffractive structure functions than 
for 
the
total ones \cite{GRV} (.4 GeV$^2$, see Fig.\ref{fig1}). 

\begin{figure}
\begin{center}
\centerline{\psfig{figure=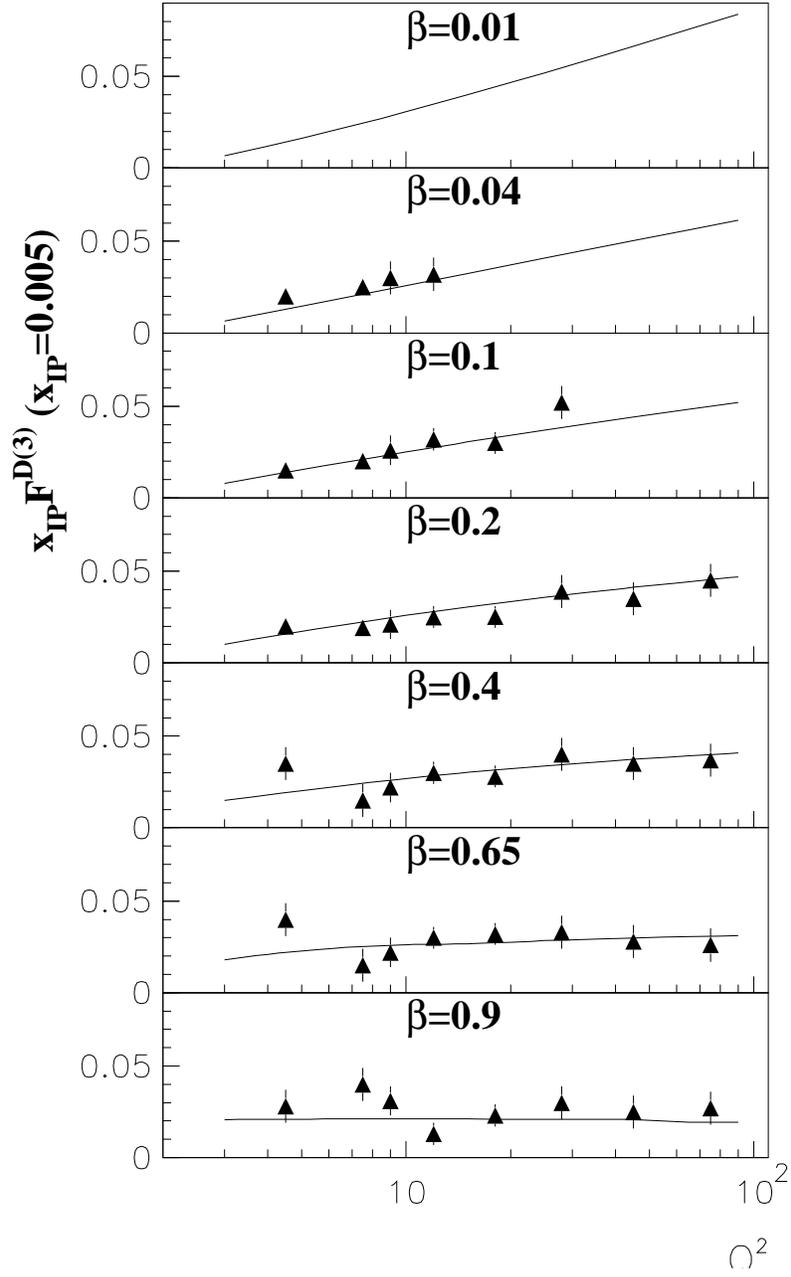,height=7.2in}}
\end{center}
\bigskip
\caption{Scaling violations. 
The prediction for the structure function
$x_{\PO} \cdot F_2^{D(3)}$ ($x_{\PO}=0.005$) is presented
as a function of $Q^2$ in
bins of $\beta$, over the full $Q^2$ range accessed.
Data points are from Ref.[14].
}
\label{laurent3b}
\end{figure}

The error band corresponds to the systematic and statistical 
errors
added in quadrature. Note that the errors are very small at low values of 
$\beta$ by 
definition
since we impose the behaviour at low $\beta$ to be proportional to $\beta$ in 
our
parametrisation, see equation (\ref{valence}). 
As shown in the figures, the result is also compatible with
the quark and gluon densities \cite{us} found using the usual H1 QCD fit using 
the input (\ref{gluon}) with the parameters of table II.
The scaling violations obtained using our parametrisation are given in Fig.
\ref{laurent3b}. We note that our parametrisation leads to positive scaling
violations for all $\beta$ values and flattens out at the highest
values of $\beta,$ in good agreement with data.

\begin{figure}
\begin{center}
\centerline{\psfig{figure=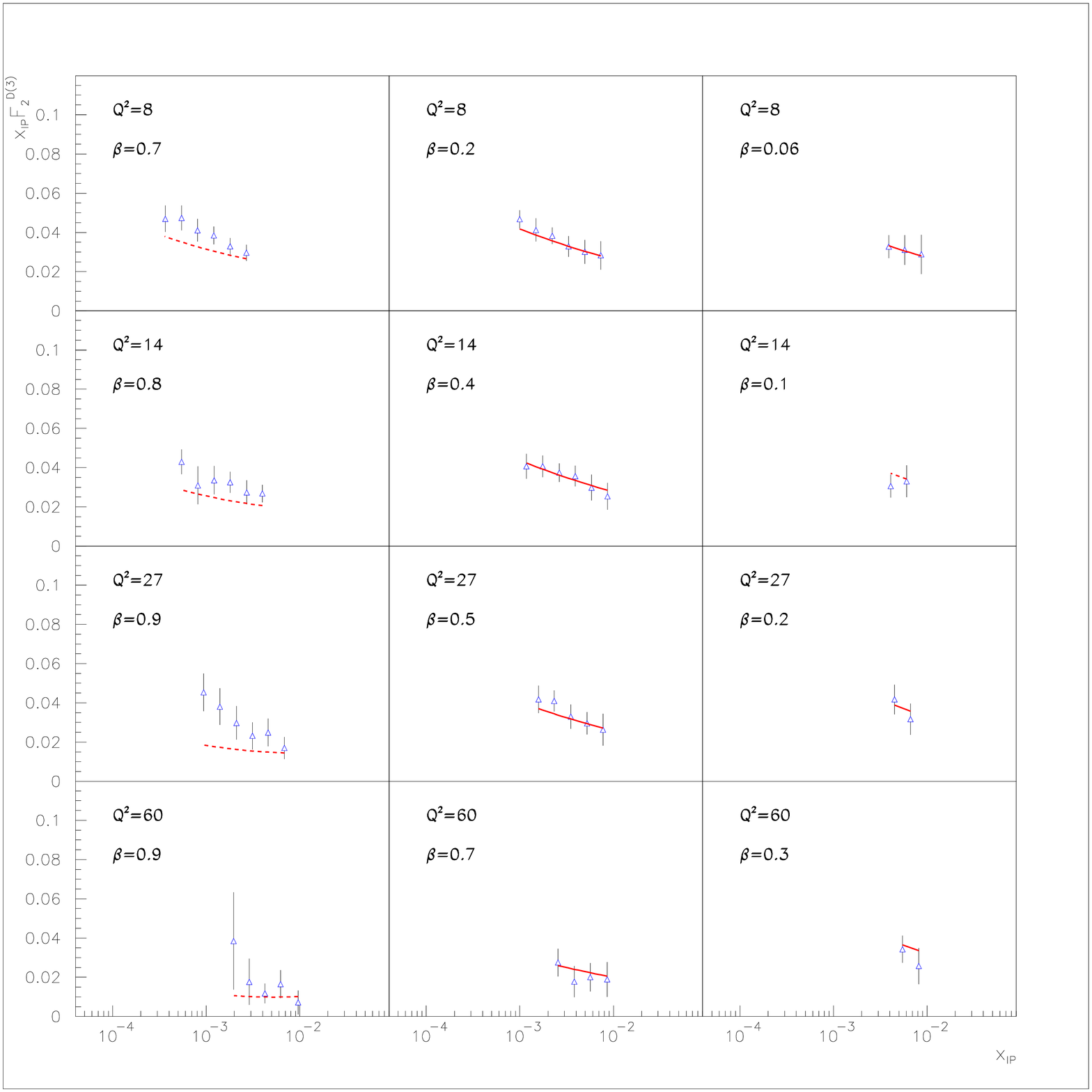,height=4.in}}
\end{center}
\bigskip
\caption{Result of the NLO QCD fit to the ZEUS $F_2^D$ data. The $\chi2$ is 
52.1 /30 data points.
We note a bad description of the data especially at high $\beta$.
}
\label{laurent5}
\end{figure}

The results for the ZEUS collaboration are given in Fig.\ref{laurent5} for the
NLO fit. 
The obtained $\chi^2$ ($\chi^2$=44.6 and $\chi^2$=52.1
respectively at LO and NLO for 30 data points) 
are two times worse than the 
one obtained by the usual QCD fit \cite{us}. We note in Fig. \ref{laurent5} that 
our
parametrisation cannot describe the data at low $\beta$. The gluon are quark 
densities
we obtain are respectively two times higher and smaller 
than 
the usual results of the fit to the ZEUS data. We are thus unable to reproduce 
ZEUS
data using our parametrisation.

\bigskip

{\bf 4. Outlook: The Nature of the Pomeron} 

\bigskip

As a brief summary of our study, the data on hard diffraction at HERA, as 
obtained from rapidity gap selection by the H1 collaboration, are well described 
using a QCD evolution of parton distributions in the Pomeron, starting mainly 
from   {\it valence-like} gluons (plus a small fraction of {\it valence-like} 
sea 
quarks) as an input at low scale. More details  have been given at the end of 
section {\bf 1}. In this last section we want to discuss the possible physical  
interpretation of this phenomenological result in terms of characteristic 
non-perturbative structures of the  enigmatic Pomeron. Note that the similar 
question about the proton was the underlying physical motivation for the GRV 
parametrizations \cite{GRV}.

Using the H1 fit results, and considering the  $j\!=\!1$ moments which are 
finite by definition of the model, it is possible to compute the number of 
quarks 
and gluons at the starting scale $\mu^2$: 
\begin{eqnarray}
&~& \frac{\int d \beta \ S}{\int \beta \ d \beta (S+G)}\  = \ 0.13\pm 0.01 \pm 
0.02 
\nonumber\\
&~& \frac{\int d \beta\  G}{\int \beta \ d \beta (S+G)}\  =\  2.1 \pm 0.1 \pm 
0.1\ . 
\label{numbers}
\end{eqnarray}
We thus find around $2.1$  for the
gluon density and around $0.1$ for the quark density, which is compatible with a 
picture of Pomeron made of two gluons
at the initial scale of  low $Q^2$.  Note that this number has nothing of an  
initial input of our study, since only  
$j=2$ moments have been  constrained by energy momentum conservation to  sum to 
$1$ at all $Q^2.$

\begin{figure}
\begin{center}
\centerline{\psfig{figure=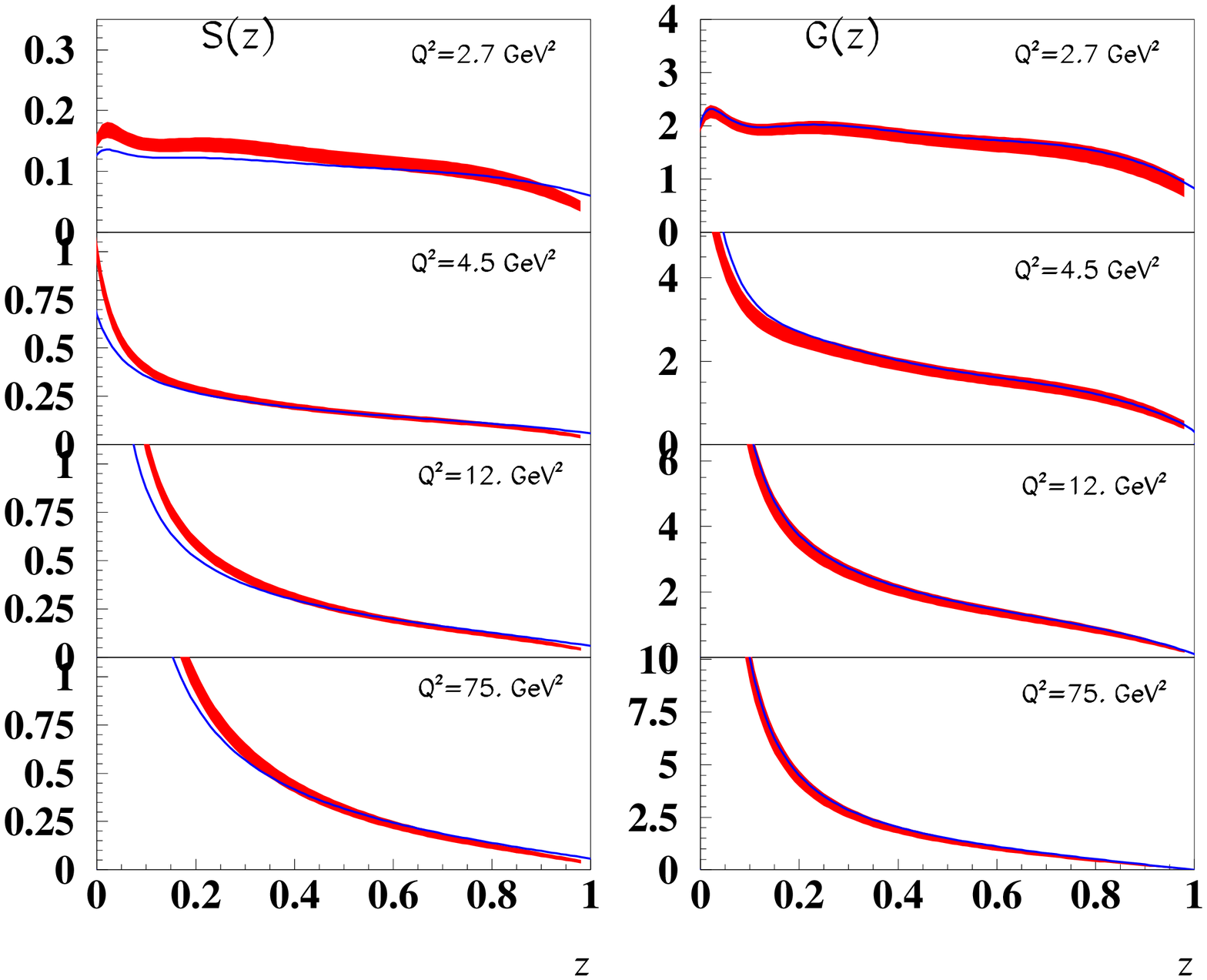,height=5.in}}
\end{center}
\bigskip
\caption{Sea quark and gluon probability densities.  The 
distributions $S(z,Q^2)$ and $G(z,Q^2)$ are displayed for the  NLO  fit to H1
data with the error bands as in Fig.\ref{laurent2}. The LO fit (mean value) is 
also shown (continuous line). The top of the figure is for the {\it 
valence-like} 
input distributions at $\mu^2.$}
\label{fig9}
\end{figure}

The probability density of gluons $G(\beta,Q^2)$ (and sea quarks 
$S(\beta,Q^2)$) gives other interesting complementary information. It is 
displayed in Fig.\ref{fig9}, for various values of $Q^2.$ It exemplifies the 
dynamical parton  generation through  QCD evolution with an increasing number of 
partons at small $\beta$ at the expense of those at high $\beta.$ In the top 
figure, we see the input probability density $G(\beta,\mu^2)$ (and sea quarks 
$S(\beta,\mu^2)$) which is almost flat in $\beta.$ This striking feature means 
that the (almost) two gluons refferred to above, have an (almost) equal 
probability in terms of energy sharing. This feature indicates a state with two 
highly interacting gluons, which cannot be interpreted as independent 
constituent gluons. This  is different   from what has been observed for the 
proton, namely a  constituent quark model for the proton or, also, from
the peaked distribution \cite{mesons} of quasi-free  heavy quarks in mesons, for 
comparison.

Summarizing the features of the obtained input distributions:
\begin{itemize}
\item  The gluon 
distribution is largely dominant over the sea quark one.
\item The total 
$j=1$ 
moments (see formula (\ref{numbers})) are near the value  2.
\item The probability densities are nearly  flat in momentum fraction.
\end{itemize}

These features are quite reminiscent of a {\it gluonium} in a fundamental 
$0^{++}$ or $2^{++}$states, where almost no $q\bar q$ excitations are present. 
Such a spectrum has been found\footnote{However a different study by Dalley and 
van de Sande \cite{glueball1} using transverse lattice technics   gives a 
spectrum decreasing at large $\beta.$} in a $1+1$-dimensional reduction of 
$SU(N)$ gauge theories.

The relatively high (compared to the proton 
case in GRV) value of the initial scale $\mu$ is also to be remarked. It is less 
stable than the other parameters, considering the variations between LO and NLO 
fits, but it stays in the range  $2.1-2.7$ GeV$^2.$ This  could be interpreted 
as 
an input Pomeron state having a rather large mass squared.  Note the 
existence of $2^{++}$ gluonium states near-by in mass \cite{glueball1}. They 
also 
have been
already discussed as possible candidates of gluonium states situated on  the 
Pomeron trajectories 
\cite{land2}.

As an outlook, It will  be interesting to look for an analytic representation of 
Pomeron QCD structure functions for all $Q^2$ in the spirit of the GRV 
parametrizations. On a more experimental ground,
it will be very interesting to verify  our conclusions with the data  announced 
by  the H1 collaboration \cite{dis2002}. A first look seems encouraging and  
these data when publicly available  will allow a better determination of the 
parameters of our input {\it valence-like} parametrization
(\ref{valence}). We thus 
expect that they will  give even more information on the elusive nature of the 
Pomeron.
\bigskip

{\bf ACKNOWLEDGMENTS}

\bigskip

 We greatly appreciated  fruitful 
discussions with S. Munier and e-mail exchanges with S. Dalley. One of us (J.L.) 
thanks the ``Service de Physique 
Th\'eorique de 
Saclay'' for  hospitality.

\eject
\vspace{1cm}

 {\bf TABLES}

\bigskip  
 
\begin{table}
\begin{center}
%\hsize=10.truecm
\begin{tabular}{|c||c|c|c|c|c|c|}
\hline  & $a_q$ & $b_q$ &$a_g$ & $b_g$ & $\mu^2$ & $\chi^2$  \\
\hline\hline
H1 LO & 0.13$\pm$0.01$\pm$0.02 & 0.40$\pm$0.02$\pm$0.13  & 
1.9$\pm$0.1$\pm$0.1  & 0.20$\pm$0.01$\pm$0.02  & 2.3 
$\pm$0.08$\pm$0.50 & 210.7/161 \\
H1 NLO & 0.13$\pm$0.01$\pm$0.01 & 0.30$\pm$0.01$\pm$0.04     & 
1.90$\pm$0.03$\pm$0.05   & 0.20$\pm$0.02$\pm$0.04   & 2.7$\pm$0.1$\pm$0.50  &  
194.6/161\\ \hline
ZEUS LO & 0.18$\pm$0.07$\pm$0.02 & 0.40$\pm$0.06$\pm$0.04  & 
2.3$\pm$0.3$\pm$0.4  & 0.02$\pm$0.02$\pm$0.01  & 0.50$\pm$0.05$\pm$0.03 & 
44.6/30 \\
ZEUS NLO & 0.18$\pm$0.02$\pm$0.02 & 0.30$\pm$0.02$\pm$0.07     & 
2.30$\pm$0.02$\pm$0.30   & 0.01$\pm$0.01$\pm$0.01   & 0.90$\pm$0.02$\pm$0.05  &  
52.1/30 \\ \hline
\end{tabular}
\bigskip
\caption{Values of the parameters obtained at LO and NLO for the fit to H1 data
(two first lines), and ZEUS data (last two lines). The first error is 
statistical
and the second systematic.
}
\end{center}
\label{I}
\end{table}

\begin{table}
\begin{center}
\hsize=8.truecm
\begin{tabular}{|c|c|c|} \hline
 parameters & H1 &  ZEUS  \\ 
\hline\hline
 $C_1^{(S)}$ &0.18 $\pm$ 0.05 &  0.41  $\pm$ 0.02 \\
 $C_2^{(S)}$ & 0.07 $\pm$ 0.02 & -0.16  $\pm$ 0.03 \\
 $C_3^{(S)}$ & -0.13     $\pm$ 0.02 & -0.11  $\pm$ 0.02 \\ \hline
 $C_1^{(G)}$ & 0.82      $\pm$ 0.40 & 0.53     $\pm$ 0.30 \\
 $C_2^{(G)}$ & 0.22 $\pm$ 0.06 & 0.28 $\pm$ 0.25 \\
 $C_3^{(G)}$ & 0.01 $\pm$ 0.04 & 0.02 $\pm$ 0.11
 \\
\hline
\end{tabular}
\vskip .5cm
\caption{Parameters for quark and gluon input distributions in the  Pomeron [5], 
see formula (\ref{gluon}).
The parameters are given at the initial scale $Q_0^2=3$ GeV$^2$ .}
\end{center}
\end{table}

 {\bf FIGURES}

\bigskip

\bigskip
{\bf REFERENCES}

\begin{thebibliography}{99}

 
       
\bibitem{bfkl}  L.N.Lipatov, {\it Sov. J. Nucl. Phys.} {\bf 23} (1976) 642;
V.S.Fadin, E.A.Kuraev and L.N.Lipatov, {\it Phys. lett.} {\bf B60} (1975)
50; E.A.Kuraev, L.N.Lipatov and V.S.Fadin, {\it Sov.Phys.JETP} {\bf 44}
(1976) 45, {\bf 45} (1977) 199; I.I.Balitsky and L.N.Lipatov, {\it %
Sov.J.Nucl.Phys.} {\bf 28} (1978) 822.


\bibitem{ingelman} G.Ingelman, P.Schlein, 
{\it Phys. Lett.} {\bf  B 152} (1985) 256.

\bibitem{f2dH1}  C.Adloff et al., H1 Col., {\it Z. Phys.} {\bf C76} (1997) 613.

\bibitem{f2dZEUS} ZEUS Col., {\it 
Eur.Phys.J.}{\bf C6 }(1999) 43.


\bibitem{us} C.Royon, L.Schoeffel, J.Bartels, H.Jung, R.Peschanski, {\it
Phys.  Rev.} {\bf D63} (2001) 074004.

\bibitem{dglap} G.Altarelli and G.Parisi,
{\it Nucl. Phys.} {\bf B126}  18C (1977) 298.
V.N.Gribov and L.N.Lipatov, {\it Sov. Journ. Nucl. Phys.} (1972) 438 and 675.
Yu.L.Dokshitzer, {\it Sov. Phys. JETP.} {\bf 46} (1977) 641.

\bibitem{GRV} M.Gl\"uck, E.Reya, A.Vogt, {\it Z. Phys.} {\bf C41} (1988) 
667, {\bf C48} (1990) 471, {\bf C53} (1992) 651, {\bf C67} (1995) 433,
{\it Eur.Phys.J.} {\bf C5} (1998) 461. Note that Fig.1 has been obtained from 
the most 
recent NLO parametrizations (1998).

\bibitem{lacaze} The NLO expressions of the  evolution kernel and coefficients 
in 
Mellin space we used can be found in E.G.Floratos, C.Kounnas and R.Lacaze,
 {\it Nucl. Phys.} {\bf B192} (1981) 417, as well as in the first reference of 
\cite 
{GRV}.

\bibitem{grad} 
I.S.Gradshteyn, I.M.Ryzhik, {\it Tables of Integrals, Series and Products}, 
Academic 
Press, San Diego, 1994.
\bibitem{mesons} S. Munier, private communication and A.C.Caldwell,  M.S.Soares 
 {\it Nucl. Phys.} {\bf A696} (2001) 4125.

\bibitem{glueball1} F.Antonuccio, S.Dalley, {\it Nucl. Phys.} {\bf B461} (1996) 
275.\\ S.Dalley, B.van de Sande,  {\it Phys.Rev.} {\bf D62} (2000) 
 014507408.
 
\bibitem{land2} P.V.Landshoff, {\it  Talk at Meeting on Elastic Scattering and 
Diffraction, Prague},  hep-ph/0108156.

\bibitem{dis2002} P.Laycock, {\it  talk given at 10th Intl. Workshop on Deep 
Inelastic Scattering (DIS 2002), Cracow, May 2002};\\
F.P.Schilling {\it  idem},  hep-ex/0209001; \\ P.R.Newman, {\it Talk at low-x 
Meeting, Antwerpen, September 2002}. 

\bibitem{royon} C.Royon, {\it Nucl. Phys. Proc. Suppl.} {\bf 79} (1999) 256.

\end{thebibliography}
\end{document}